\documentclass[aps,prb,twocolumn,superscriptaddress]{revtex4-1}

\usepackage{hyperref} % hyperreferences
\usepackage{graphicx}
\usepackage{amsmath}
\usepackage{amssymb} 
\usepackage{hyperref}
\usepackage[dvipsnames,svgnames,x11names,hyperref]{xcolor}
\hypersetup{ colorlinks=true,linkcolor=NavyBlue,urlcolor=NavyBlue,  citecolor=NavyBlue}
\DeclareGraphicsExtensions{.png,.jpg,.eps}
\usepackage{xcolor}

\begin{document}

\author{Alejandro Lopez-Bezanilla}
\affiliation{Theoretical Division, Los Alamos National Laboratory, Los Alamos, New Mexico 87545, United States}

\author{J. L. Lado}
\affiliation{Department of Applied Physics, Aalto University, 00076 Aalto, Espoo, Finland}

\title{Electrical band flattening, valley flux,  
and superconductivity
in twisted trilayer graphene
}

\begin{abstract}
	Twisted graphene multilayers have demonstrated to yield a versatile playground to engineer con-
trollable electronic states. Here, by combining first-principles calculations and low-energy models,
we demonstrate
	that twisted graphene trilayers provide a tunable system where van Hove singularities can
be controlled electrically. 
	In particular, it is shown 
	that besides the band flattening, bulk valley currents
appear, which can be quenched by local chemical dopants. We finally show 
	that in the presence of
electronic interactions, a non-uniform superfluid density emerges, whose non-uniformity gives rise to
spectroscopic signatures in dispersive higher energy bands. Our results put forward twisted trilayers
as a tunable van der Waals heterostructure displaying electrically controllable flat bands and bulk
valley currents.
\end{abstract}

\date{\today}

\maketitle

%%%%%%%%%%%%%%%%%%%%%%%%%%%%%%%%%%%%%%%%%%%%%%%%
%%%%%%%%%%%%%%%%%%%%%%%%%%%%%%%%%%%%%%%%%%%%%%%%
%%%%%%%%%%%%%%%%%%%%%%%%%%%%%%%%%%%%%%%%%%%%%%%%

\section{Introduction}
The interplay between topology and correlations represents
a highly fruitful area in condensed matter physics.
However,
exploring unconventional states of matter requires
identifying systems where electronic correlations, topology
and electronic dispersions can be realistically controlled.
In this line, twisted van der Waals materials\cite{Cao2018super,Yankowitz2019,Liu2014,Lu2019,Rickhaus2018,PhysRevB.88.121408,Liao2020,Shimazaki2020}
provide
a powerful solid state platform to realize exotic quantum phenomena.
The tunability
of twisted van der Waals materials
stems from the emergence of a band structure that
can be controlled
by the twist between different two-dimensional materials.\cite{PhysRevLett.99.256802,PhysRevB.82.121407}
In particular,
the different quantum states in twisted graphene systems stem
from the possibility of controlling the ratio between kinetic and
interaction terms. In twisted graphene bilayers,
such tunability allowed to realize
superconducting,\cite{Cao2018super,Yankowitz2019,Lu2019} correlated
insulators, topological networks,\cite{Rickhaus2018,PhysRevB.88.121408}
Chern insulators\cite{Serlin2019}
and quasicrystals.\cite{Ahn2018,PhysRevB.99.165430,Yu2019,2020arXiv200110427P}
As a result, current experimental efforts are focusing on
exploring new twisted van der Waals materials,
with the aim of finding platforms that allow for an
even higher degree of control.\cite{2020arXiv200411340C,2020arXiv200411353P}

From the quantum engineering point of view,
applying a perpendicular bias between layers\cite{PhysRevLett.99.216802} provides a versatile
way of tuning correlated states in twisted graphene multilayers.
This has been demonstrated
in paradigmatic examples of correlated states in
twisted tetralayers (double bilayer)\cite{2019arXiv190308130L,Shen2020} 
and twisted trilayers
(monolayer/bilayer).\cite{2020arXiv200411340C,2020arXiv200411353P}
Moreover, interlayer bias is known to generate internal
valley currents in twisted graphene bilayers,\cite{Rickhaus2018,PhysRevB.88.121408,PhysRevLett.121.146801,PhysRevLett.123.096802}
creating
topological networks at low angles\cite{Rickhaus2018,PhysRevB.88.121408,PhysRevLett.121.146801} and
generating valley fluxes in flat bands regimes.\cite{PhysRevLett.123.096802}
This interplay of correlations and topology in twisted graphene
multilayers makes these materials a powerful platform to explore
exotic states of matter\cite{PhysRevLett.124.106803,2020arXiv200409522L,2019arXiv191211469R,2019arXiv191209634L}
in a realistically feasible manner.

From the theoretical point of view,
electronic structure calculation of
twisted graphene bilayer conducted with
real-space tight-binding models\cite{PhysRevB.82.121407} or continuum Dirac descriptions\cite{PhysRevLett.99.256802}
capture
the fundamental features of the electronic dispersion.
Nevertheless, internal coordinates optimization can quantitatively modify the electronic
dispersion.\cite{PhysRevB.98.224102,PhysRevB.98.195432,PhysRevB.96.075311,2019arXiv191012805L,Brihuega2017,PhysRevB.100.075416,PhysRevX.9.041010,Rickhaus2019} Well known examples of this are the growth of AB/BA regions
in twisted bilayers.\cite{PhysRevB.96.075311,PhysRevB.100.155426}
It is important to note that studying twisted graphene multilayers
from first-principles represents a remarkable
challenge, due to the large amount of atoms
present in a unit cell.

\begin{figure}[t!]
\centering
    \includegraphics[width=\columnwidth]{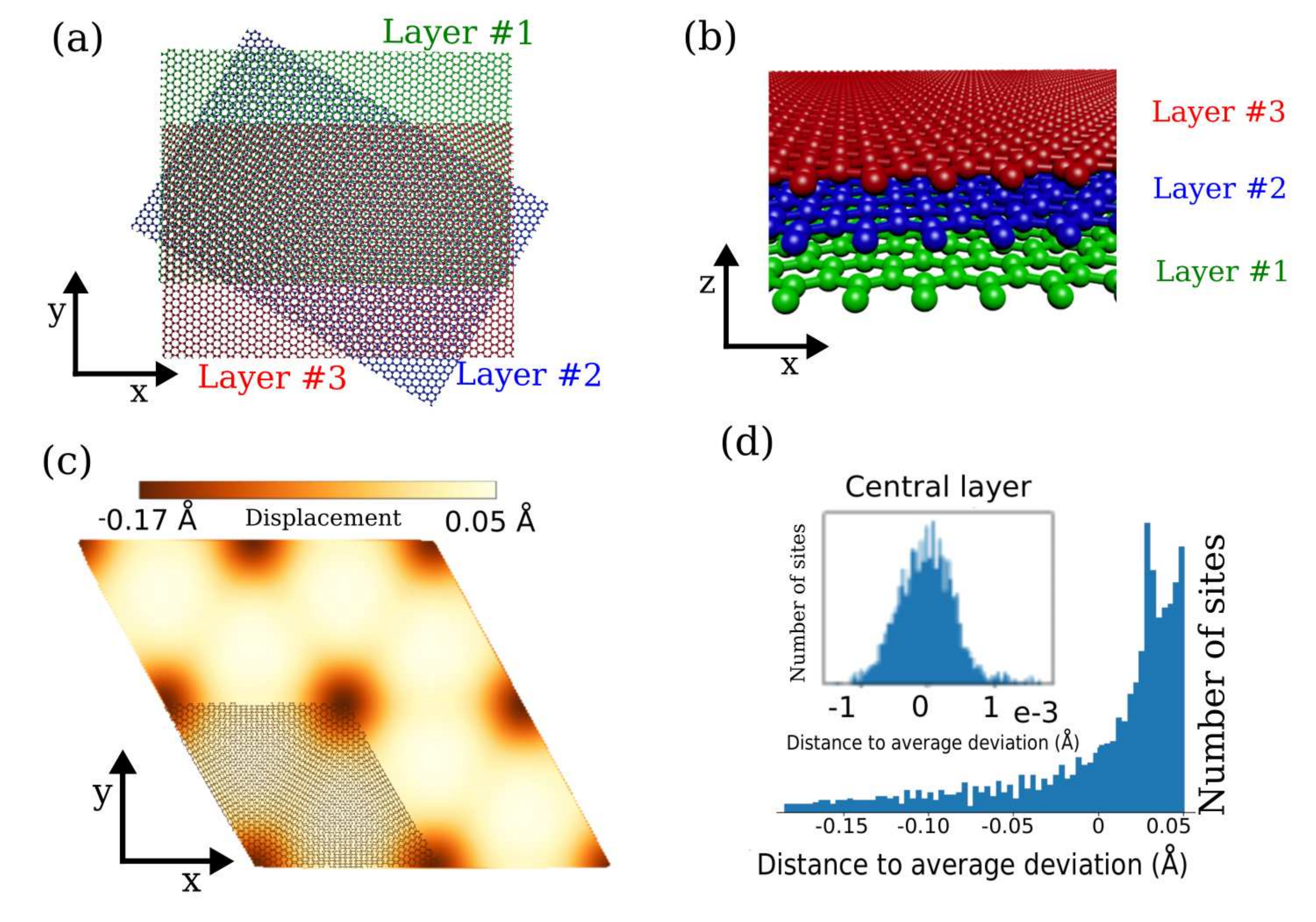}

\caption{(a,b) Sketch of the structure a twisted graphene trilayer,
in which the top and bottom layers are aligned, and twisted with respect the middle
one.
Panel (c) shows the color map of a DFT-based relaxed 2$\times$2 supercell of twisted trilayer. Colored pattern corresponds to the difference $\Delta = z-z_{av}$ between the z coordinate of an atom in surface layer and the average coordinate $z_{av}$ of all atoms in the same layer. AA stacking regions are colored with darker tones whereas AB/BA stacking regions are colored with lighter tones. Superimposed is the unit cell of the trilayer. 
Panel (d) shows the histogram shows the number of occurrences of each distance to the average deviation value. Tiny deviations from flatness occur in the central layer.
Calculations performed with DFT.
}
\label{figureContour}
\label{fig:sketch}
\end{figure}

Here, by combining first-principles calculations and low-energy models, we
show that twisted graphene trilayers\cite{PhysRevB.87.125414}
host flat bands whose bandwidth
can be controlled electrically. We address the 
impact of an interlayer bias both from first-principles
and effective models, showing that van Hove singularities can be
merged electrically.
We show that associated with the interlayer bias, bulk valley currents emerge,
that are impacted by the existence of chemical impurities in the system.
We finally address the superconducting states in these doped trilayers,
showing that the non-uniform superfluid density has an impact
in high-energy bands. Our manuscript is organized as follows,
in Sec. \ref{sec:bands} we show the electronic structure
of twisted graphene trilayer both from first-principles
and low-energy models, in Sec. \ref{sec:field} we explore
in detail the impact of an electric field,
in Sec. \ref{sec:imp} we explore the effect of chemical impurities,
and
in Sec. \ref{sec:sc} we address
the impact of an emergent non-uniform superfluid density. Finally,
in Sec. \ref{sec:sum} we summarize our conclusions.

\section{Electronic structure of twisted trilayer graphene}
\label{sec:bands}

The electronic structure of twisted graphene trilayers\cite{PhysRevB.87.125414,2019arXiv190712338L,PhysRevB.100.085109,2019arXiv190700952C,PhysRevLett.123.026402,2020arXiv200501258P} shows
different features in comparison
with twisted graphene bilayers.\cite{PhysRevB.82.121407,Bistritzer2011}
The electronic structure of small "magic" angle tBLG
features four flat bands lying around the Fermi level, with 
band splitting at the $\Gamma$ point of the Brillouin zone of
the emergent moir{\'e} superlattice.\cite{PhysRevB.82.121407,Bistritzer2011}
Density functional theory (DFT) calculations have shown
consistent results with those effective models
in twisted graphene multilayers, yet
quantitative modifications are observed when including
relaxation of the atomic coordinates\cite{PhysRevB.96.075311,PhysRevB.98.235137,2019arXiv191012805L,PhysRevB.99.195419,2020arXiv200414323C}
and crystal-field effects\cite{Rickhaus2019,Haddadi2020}.
Therefore, to benchmark the electronic properties 
of twisted graphene multilayers, it is essential to start from a 
correct description that takes into account the 
geometric corrugation and ab-initio
electrostatics of the moir{\'e} system.

We consider a twisted trilayer structure in which
the upper and lower layers are aligned,
and the middle one is twisted with an angle $\theta$ with respect to those (Fig. \ref{fig:sketch}ab).
In particular, in the following we will consider a twisted trilayer whose middle layer has
a twisting angle of 1.9$^\circ$ with respect to the external layers. 
A twisted multilayer like this can be created with standard tear, rotate and stack
techniques.\cite{Kim2016rotate}
The system of 5514 atoms 
is fully relaxed allowing for lateral and vertical displacement of the C atoms in the structure.
Figure \ref{figureContour}c shows the color map of one of the two equivalent external layer relaxation. The color scheme
represents the vertical variation of each C atom at the surface with respect to the average deviation within each
layer. The darker areas indicate a displacement of atoms out of the surface
which occurs predominantly in the AA stacking
region. The histogram of Fig. \ref{figureContour}d shows 
that the number of atoms in the upper layer whose vertical coordinate is above the 
average is twice as large as the atoms displaced in the opposite direction
below the average.

\begin{figure}[t!]
\centering
\includegraphics[width=\columnwidth]{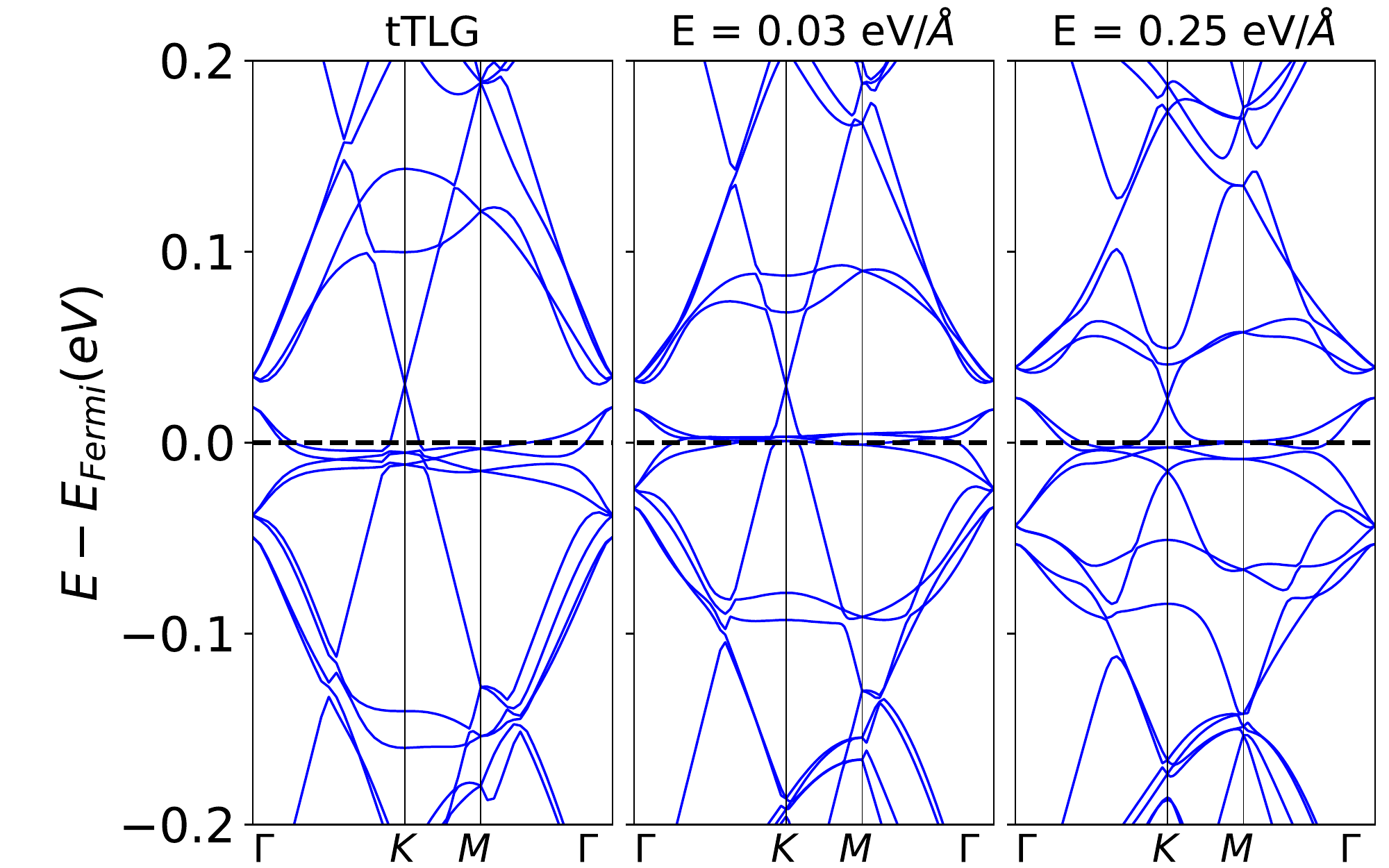}

\caption{Electronic band diagrams of twisted trilayer graphene. Localized states at the Fermi energy coexist with highly dispersive states bands forming a Dirac cone in the conduction band. Applying a small electric field perpendicular to the TTG a reduction of all bands dispersion is induced. A stronger electric field removes the Dirac cones and induces hybridization between electronic states.  
Calculations performed with DFT}
\label{figEfields}
\label{fig:dftf}
\end{figure}

The first-principles
electronic band diagram of the fully relaxed structure of 
$\theta=1.9^\circ$ TTG
is shown in Fig. \ref{figEfields}.
In contrast with twisted bilayer graphene, two 
highly dispersive bands coexist with 
four low-dispersive bands grouped at the Fermi energy. 
The Dirac-like crossing above the charge neutrality point
leads to a small charge transfer between the flat and dispersive bands
even at half filling.
Additional flattening of the localized states can be induced by application of an external 
electric field 
perpendicular to the TTG surface. A field of 0.03 eV/\AA\ reduces 
the dispersion of all electronic states in the vicinity of 
the Fermi energy without inducing any inter-band charge transfer. 
Increasing the strength up to 0.25 eV/\AA\ a disruption 
of the linear bands is observed and a hybridization of the flat bands 
with neighboring state increase their dispersion. It is observed that 
the application of an interlayer bias generates Dirac crossings above and
below charge neutrality, besides a variety of anticrossings
(Fig. \ref{figEfields}).

The first-principles calculations above show that 
the electronic structure of twisted trilayer graphene show strong differences
with the one of twisted graphene bilayers. 
In particular,
a highly dispersive set of
bands coexists
with the nearly
flat bands at charge neutrality. In order to explore more in detail
the physics and twisted trilayers, in the following we will
exploit a low-energy model.
We find that the tight binding model qualitatively reproduces
the important features of the band structure without including
relaxations and additional charge transfer effects. 
Therefore, for the sake of simplicity
we now take an unrelaxed structure for our tight binding calculations.
We take a single orbital per carbon atom, yielding
a tight-binding Hamiltonian of the form

\def\br{{\bold{r}}}

\begin{eqnarray}
        \mathcal{H}_0 = -t \sum_{\langle i, j \rangle,s} c_{i,s}^\dagger c_{j,s} -
        \sum_{i,j,s} \bar t_{\perp} (\br_i,\br_j) c_{i,s}^\dagger c_{j,s},
\end{eqnarray}
with
$
\bar t_{\perp}(\bold r_i,\bold r_j) =
t_{\perp}
\frac{(z_i - z_j)^2 }{|\bold r_i - \bold r_j|^2}
e^{-\beta (|\bold r_i - \bold r_j|-d)}$, where
$d$ is the interlayer distance and $\beta$
controls the decay of the interlayer hopping.
As a reference, for twisted graphene multialyers
$t \approx 3$ eV and $t_\perp \approx 0.15 t$.
\footnote{At low energies the spectra is invariant upon rescaling
of the interlayer coupling, which allows to explore effective smaller angles
with smaller unit cells.\cite{PhysRevB.98.195101,PhysRevLett.119.107201}
Our calculations are performed with a resscaled $t_\perp = 0.4 t$.}
Similar real-space models were used to study
a variety of twisted graphene multilayers,\cite{PhysRevB.82.121407,PhysRevB.87.125414,PhysRevB.82.121407,2019arXiv191101347C,PhysRevB.91.035441} 
providing a simple formalism to study the effect of dopands
and impurities.\cite{PhysRevB.99.245118,PhysRevMaterials.3.084003} 
However, in contrast
to continuum models,\cite{PhysRevB.86.155449,PhysRevLett.99.256802,Bistritzer2011} 
measuring of valley related quantities
with a real-space based formalism is non-trivial.

Twisted graphene multilayers own an approximate
symmetry associated to the valley quantum number.\cite{PhysRevLett.99.256802} 
Valley physics in both
DFT calculations and tight-binding are emergent
symmetries, in the sense that valley are not
easily defined in terms of real-space chemical orbitals.
This limitation can be overcome by defining the so-called
valley operator\cite{PhysRevLett.120.086603,PhysRevLett.121.146801,PhysRevB.99.245118}
in the tight-binding description. 
With the valley operator 
the expectation value of the valley can be computed in a real
space representation.\cite{Wolf2019,2020arXiv200305163M} 
The details of the 
valley operator\cite{PhysRevLett.120.086603,PhysRevLett.121.146801,PhysRevB.99.245118} 
are given in Sec. \ref{sec:valley}, in the following
we will take as starting point the valley operator $\mathcal{V}_z$
With the previous operator, we can compute the valley flavor of each
eigenstate of the twisted trilayer supercell
within the real-space formalism as
$\langle \mathcal{V}_z \rangle = \langle \Psi | \mathcal{V}_z | \Psi \rangle | $. 
It is worth noting that
this operator can be easily defined in the tight-binding
basis but not in the DFT basis. We finally note that the
valley operator $\mathcal{V}_z$ in the twisted moire system
will show the valley flavor in the original Brillouin zone
of graphene, not the mini-Brillouin zone of the twisted system.

\begin{figure}[t!]
\centering
    \includegraphics[width=\columnwidth]{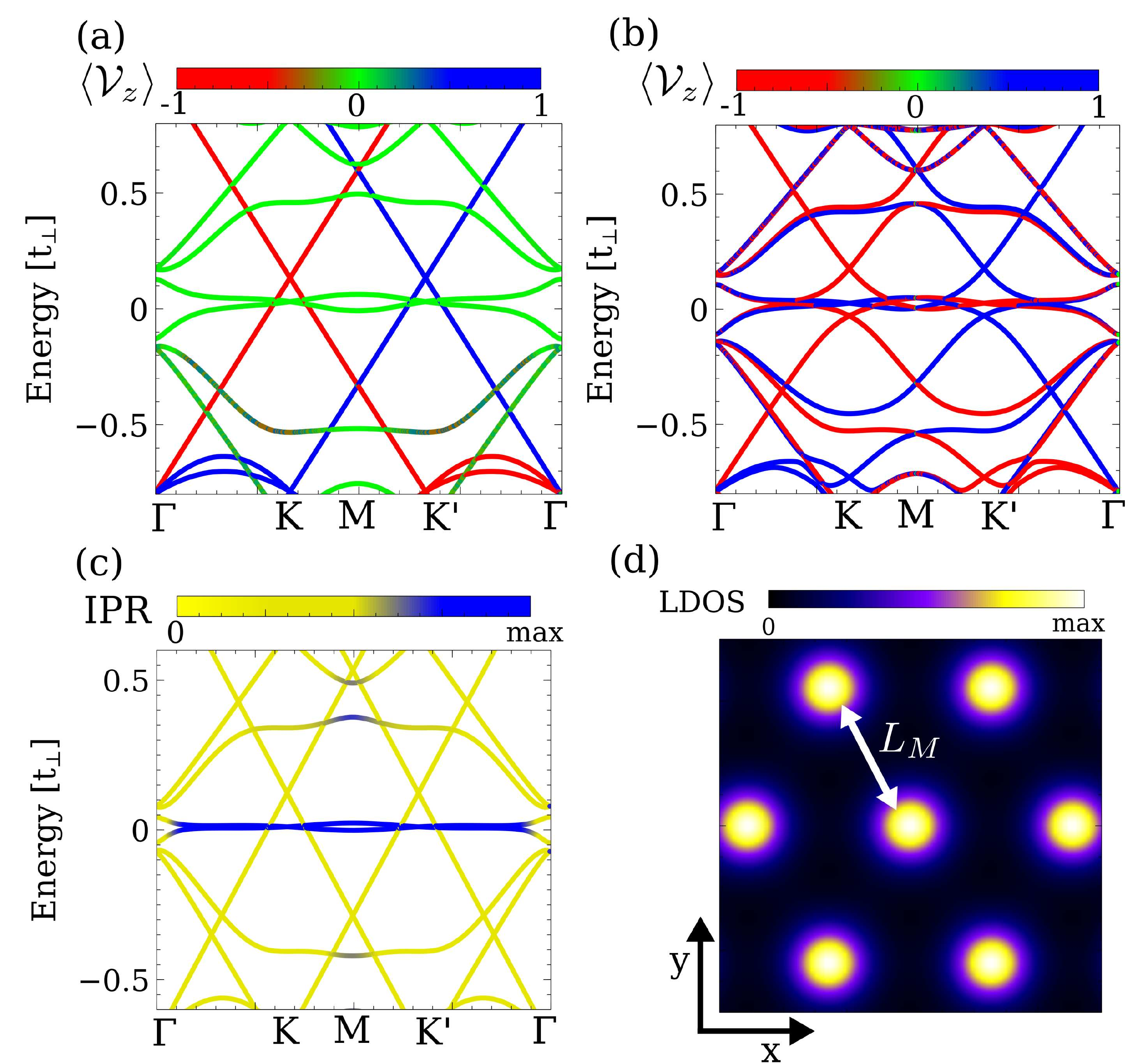}

\caption{
(a) Band structure of a twisted trilayer at 
angle of 1.9$^\circ$, without (a) and with (b) interlayer bias,
showing analogous dispersion to those found by first
principles in Fig. \ref{fig:dftf}.
Panel (c) shows the band structure
at a lower angle of
$\theta = 1.6^\circ$, showing a flattening of the bands.
Panel (d) shows the local density of states at the Fermi energy,
showing a triangular supercell pattern.
Note that part of the k-path is different
in comparison with Fig. \ref{fig:dftf}.
Calculations performed with tight-binding.
}
\label{fig:tbnof}
\end{figure}

With the previous formalism, we now compute the electronic structure
of the low-energy model at the same angle $\theta = 1.9^\circ$ as
in the first-principles calculations (Fig. \ref{fig:tbnof}).
We note that, besides some additional splittings observed
in the first-principles calculations
(Fig. \ref{fig:dftf}), the band-structure
obtained with the low-energy model (Fig. \ref{fig:tbnof}ab)
gives comparable results.
As shown in Fig. \ref{fig:tbnof}a, in the absence of an
interlayer bias the system shows nearly flat bands
coexisting with highly dispersive states. Whereas the
nearly flat bands are degenerate in valleys in the
path shown, the dispersive states belong to different
valleys in different parts of the Brillouin zone.
It is also observed that the dispersive 
Dirac cones are slightly displaced
from charge neutrality, as observed in the first
principles results.

We now move on to considered the effect of the itnerlayer bias.
An interlayer bias can be easily
included in the low-energy
tight-binding model by means of

\begin{equation}
    \mathcal{H}_V = V/(2d)\sum_{i,s} z_i c^\dagger_{i,s} c_{i,s}
\end{equation}
where $z_i$ is the $z$-position of site $i$, $d$ is the interlayer distance 
and $V$ is the strength of the interlayer bias. The full Hamiltonian
in the presence of interlayer bias is thus $\mathcal{H} = \mathcal{H}_0 + \mathcal{H}_V$.
When an interlayer bias is turned on (Fig. \ref{fig:tbnof}b),
the valley degeneracy in the $G-K$ path is lifted, and the
dispersive bands show two crossings above and below the
nearly flat bands, analogously to the
first-principles results. It is also observed that the nearly
flat bands are slightly modified by the interlayer bias.
As we will show below, such effect becomes stronger for even smaller
angles.

A last interesting point is related with the 
the localization of the different states in the supercell.
This can be characterized by means
of the inverse participation ratio (IPR), defined as
$IPR = \sum_i |\Psi_k (i)|^4$. Large values of the IPR correspond to states localized in the
moir{\'e} supercell, whereas low values correspond to delocalized in the moir{\'e}
supercell. 
Focusing now in a structure with
$\theta = 1.6^\circ$ twisting angle,
it is clearly observed that the nearly flat bands show a substantially
higher degree of localization than
the dispersive bands (Fig. \ref{fig:tbnof}c).
In particular, the flat band states are associated to an emergent
triangular lattice in the supercell, as shown in the local
density of states (LDOS) of Fig. \ref{fig:tbnof}d. In the
following we will see how these flat band states can
be electrically controlled, and how the bias
creates bulk valley currents associated with the valley
splittings.

\section{Band flattening and valley currents by an interlayer bias}
\label{sec:field}

\begin{figure}[t!]
\centering
    \includegraphics[width=\columnwidth]{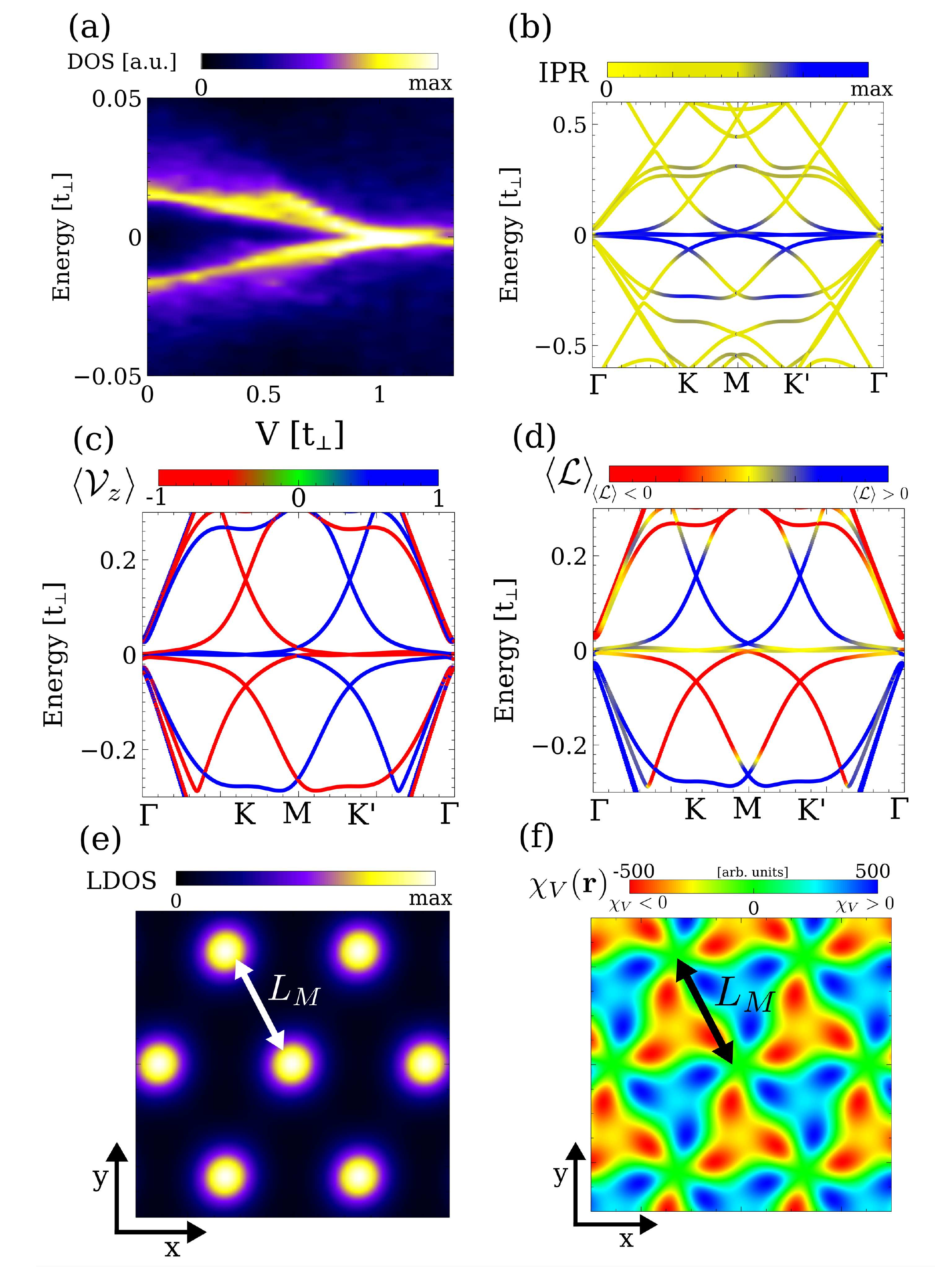}

\caption{
(a) Evolution of the density of states as a function of the interlayer bias
	for $\theta = 1.6^\circ$.
Panel (b) shows the band structure at a finite electric field depicting the
localization of each state of each state.
Panel (c) shows the valley polarization of the low-energy states, whereas
panel (d) their layer polarization.
Panel (e) shows the spatial localization of the nearly flat bands.
Panel (f) shows the local valley flux at one electron per unit cell.
The spatial maps in (e,f) are summed over the three layers.
Calculations performed with tight-binding.
}
\label{fig:tbf}
\end{figure}

We now move on to systematically
analyze the effect of an interlayer bias in the
twisted graphene trilayer, and in particular its effect on the low
energy density of states.
Electric biases in twisted graphene multilayers are known to give rise
to valley Hall currents,\cite{PhysRevB.74.161403,PhysRevLett.100.036804,Zhang2013,PhysRevB.75.155424,PhysRevB.88.121408,PhysRevLett.121.146801,Rickhaus2018}
and represent an effective knob to control
the low-energy electronic structure.\cite{PhysRevLett.99.216802} 
In the following we focus on the structure at $\theta = 1.6^\circ$.
As shown in Fig. \ref{fig:tbf}a, the application of an interlayer
bias merges the original two van Hove singularities of zero bias in
a single one, dramatically enhancing the low-energy density of states. 
Interestingly,
this merging of van Hove singularities is similar to the evolution with the twist angle in
twisted 
bilayers,\cite{PhysRevB.82.121407,Bistritzer2011,PhysRevLett.108.216802} 
with the key difference
that in the present case the merging is electrical,
turning the process highly controllable in-situ.

We now focus on the flattest regime with finite bias,
when the two van Hove singularities get merged.
Fig. \ref{fig:tbf}bcd shows the band structure of that regime .
It is observed that almost perfectly flat bands are present at the
Fermi energy (Fig. \ref{fig:tbf}bcd), accounting for the
van Hove singularity at charge neutrality (Fig. \ref{fig:tbf}a).
It is also observed that the states remain highly localized
in the unit cell as highlighted by the IPR, whereas at higher energies the
dispersive bands are delocalized in the moir{\'e} unit cell (Fig.
\ref{fig:tbf}b).
It is also observed that the bands show perfect valley polarization
(Fig. \ref{fig:tbf}c), so that the interlayer bias
does create any intervalley scattering in agreement
with continuum models. Given that the
interlayer bias breaks the symmetry between the top
and bottom layers, it is interesting to look at the
layer polarization of the states $\langle \mathcal{L} \rangle$,
defined by the layer polarization operator
$\mathcal{L} = \sum_{i,s} z_i c^\dagger_{i,s} c_{i,s}$,
with $z_i$ the $z$-component of the site $i$.
As shown in Fig. \ref{fig:tbf}d, the Dirac cones
at the $K,K'$ points
above and below the flat bands have a slightly
opposite layer polarization. However, such
layer polarization is reversed at the $\Gamma$ point (Fig. \ref{fig:tbf}d),
highlighting that the states remain highly entangled between
all the layers. 

We now
explore how the band flattening is accompanied with the emergence of bulk valley currents.
The emergence of valley currents associated with interlayer biases
is a well known effect in aligned graphene
bilayers\cite{PhysRevB.74.161403,PhysRevLett.100.036804,Zhang2013,PhysRevB.75.155424}
and tiny-angle twisted bilayers.\cite{PhysRevB.88.121408,PhysRevLett.121.146801,Rickhaus2018}
Such valley currents arise due to the emergence of valley
a non-zero local valley Chern numbers,\cite{Ren2016} whose quantization
is associated to the emergent valley conservation.\cite{RevModPhys.81.109}
In twisted systems, the local Chern number is expected to change
from region to region,\cite{PhysRevB.88.121408} due to the locally
modulated Hamiltonian. To address this local
topological property, in the following
we compute the local 
valley flux by means of the Berry flux density $\chi_V$.
The real-space valley flux $\chi_V$ defines the valley Chern number as
$C_V = C_K - C_{K'} = \int \chi_V(\mathbf r,\omega) d^2 \mathbf r d\omega$.
The real-space valley flux $\chi_V$ can be computed as\cite{PhysRevB.84.205137,Wolf2019,2020arXiv200305163M}

\begin{align}
    \chi_V(\mathbf r,\omega) = \int \frac{d^2 \mathbf k}{(2\pi)^2} \frac{\epsilon_{\alpha \beta}}{2}
 \langle \mathbf r  | G_V (\partial_{k_\alpha}G_V^{-1}) (\partial_{k_\beta}G_V)| \mathbf r \rangle
\end{align}
where $\epsilon_{\alpha\beta}$ denotes the Levi-Civita tensor, $G_V$ the 
valley Green's function $G_V = [\omega - \mathcal{H}_{\mathbf k}+i0^+]^{-1} 
\mathcal{V}_z$,
$\mathcal{H}_k$ the Bloch Hamiltonian, 
and $\mathcal{V}_z$ the valley polarization operator
of Eq. \ref{eq:valley}.

Fig. \ref{fig:tbf}f shows the spatial profile of the valley flux density
at the Fermi energy $\chi_V(\mathbf r) \equiv \chi_V(\mathbf r,\omega=\epsilon_F)$,
for a chemical potential with one hole per unit cell. It can 
be observed clearly that sizable valley currents
appear in regions in the complementary regions to the low-energy
states, similarly to other moir{\'e} systems.\cite{Wolf2019,2020arXiv200305163M}. 
Since the emergence of such currents relies on valley conservation,
terms in the system creating intervalley mixing are expected
to substantially impact them,\cite{PhysRevB.88.121408,PhysRevB.86.205414,PhysRevB.93.041407,PhysRevB.94.134201} 
as we address in the next section.

\section{Impact of chemical doping}
\label{sec:imp}

In this section we address the modification of the electronic structure under both electrical and
chemical doping. First, we start with the study of the electric doping from first-principles, and
afterwards we move on to consider its effect in the low-energy effective model.

\begin{figure}[t!]
\centering
\includegraphics[width=\columnwidth]{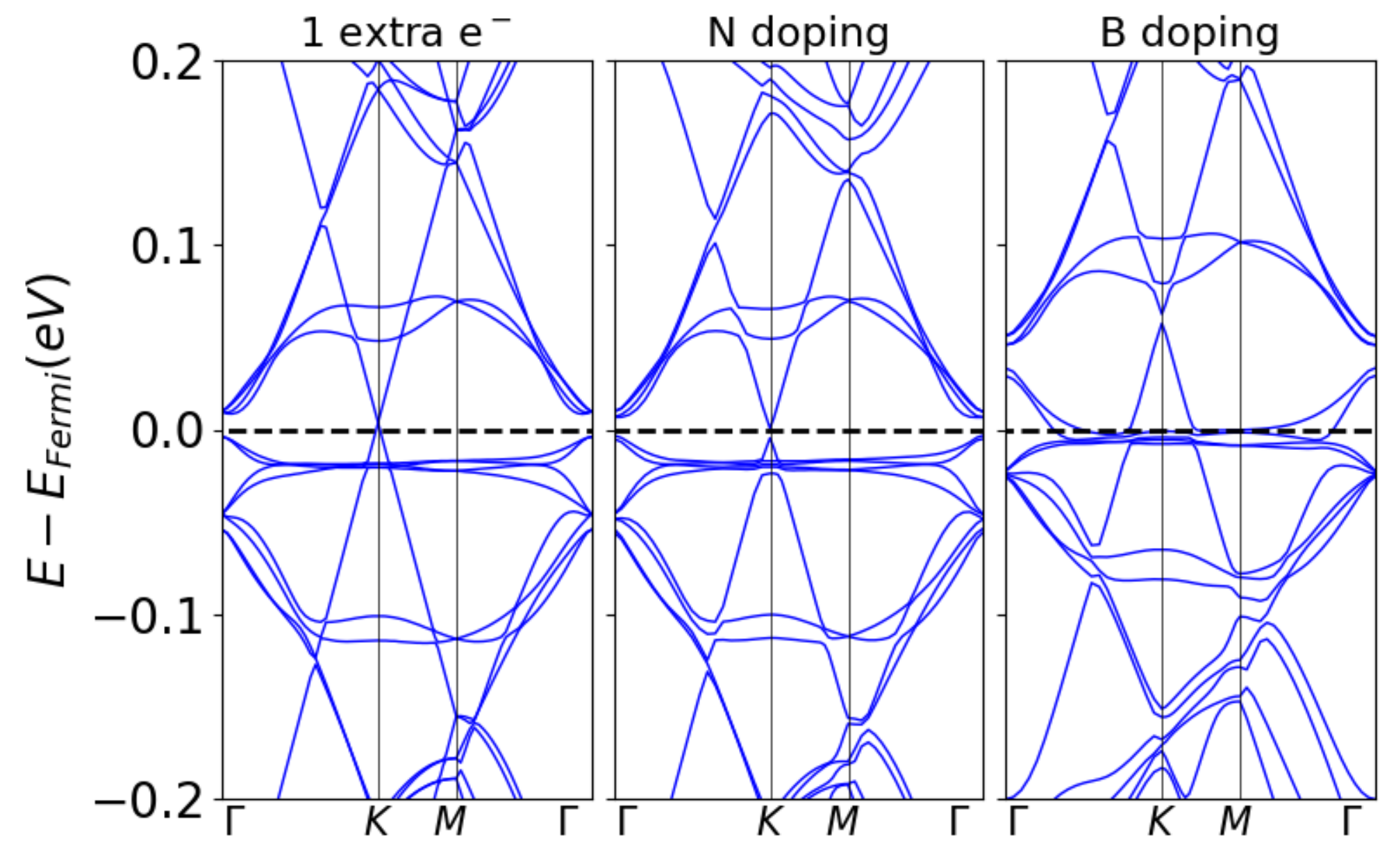}

\caption{Electronic band diagrams of three doped TTG configurations. Left panel shows 
electrostatic doping with one extra electron. Central panel shows
that doping with a   N atom causes a similar effect but 
removes the Dirac point and converts TTG in a tiny-gap semiconductor. 
Right panel shows that doping with a  
B atom empties some bands at the Fermi level including the highly dispersive bands.
Calculations performed with DFT.
}
\label{figDoping}
\label{fig:dopdft}
\end{figure}

We first address the impact of doping from first-principles.
First-principles DFT calculations are employed to study the modification of the 
band diagram of TTG upon doping with both an extra charge and one of the layers with foreign species.
Uniform doping of the TTG is realized by introducing an extra electron which is compensated with an 
equally uniform background charge of opposite sign. 
This can be realized experimentally
via field effect. 
Figure \ref{fig:dopdft} shows that the extra electron fills the flat states and align the Dirac point 
with the Fermi energy. Small modifications in the electronic band
structure as a result of new  electrostatic contributions is observed, similarly to other twisted multilayer
systems.\cite{PhysRevB.100.205113,Guinea2018,2020arXiv200305050P,2020arXiv200414784G}
 
A similar effect can be induced with chemical doping, namely adding one N atom in substitution of a C atom. 
The effects of chemical substitution have been extensively studied in graphene,\cite{PhysRevB.93.245420,PhysRevB.83.245403,Liu2011,LopezBezanilla2009,Agnoli2016,Georgakilas2012,PhysRevLett.109.226802,PhysRevLett.104.187201,LopezBezanilla2009_2,PhysRevLett.97.236802}
including twisted bilayers.\cite{PhysRevMaterials.3.084003,2019arXiv190809555Y}
Figure \ref{fig:dopdft} shows the effect on the bands structure of one N atom in one of the surface layers. 
The extra charge supplied the N atom shifts the chemical potential, increasing
the filling of the flat bands. Major 
difference with respect to the electrostatic doping is the opening 
of a meV large band gap as a result of the 
symmetry breaking imposed by the impurity which 
also induces a mixing of the linear states with the less 
dispersive states. These new anticrossings are 
due to the intervalley scattering created by the chemical 
impurity, and they do not appear
in the case of electrostatic doping.
A similar hybridization of electronic states is also 
observed when one electron is removed by means of 
doping with a B atom, as shown in Fig. \ref{fig:dopdft}. The chemical potential shift occurs in the opposite 
direction and the filling of flat band states decreases.

\begin{figure}[t!]
\centering
    \includegraphics[width=\columnwidth]{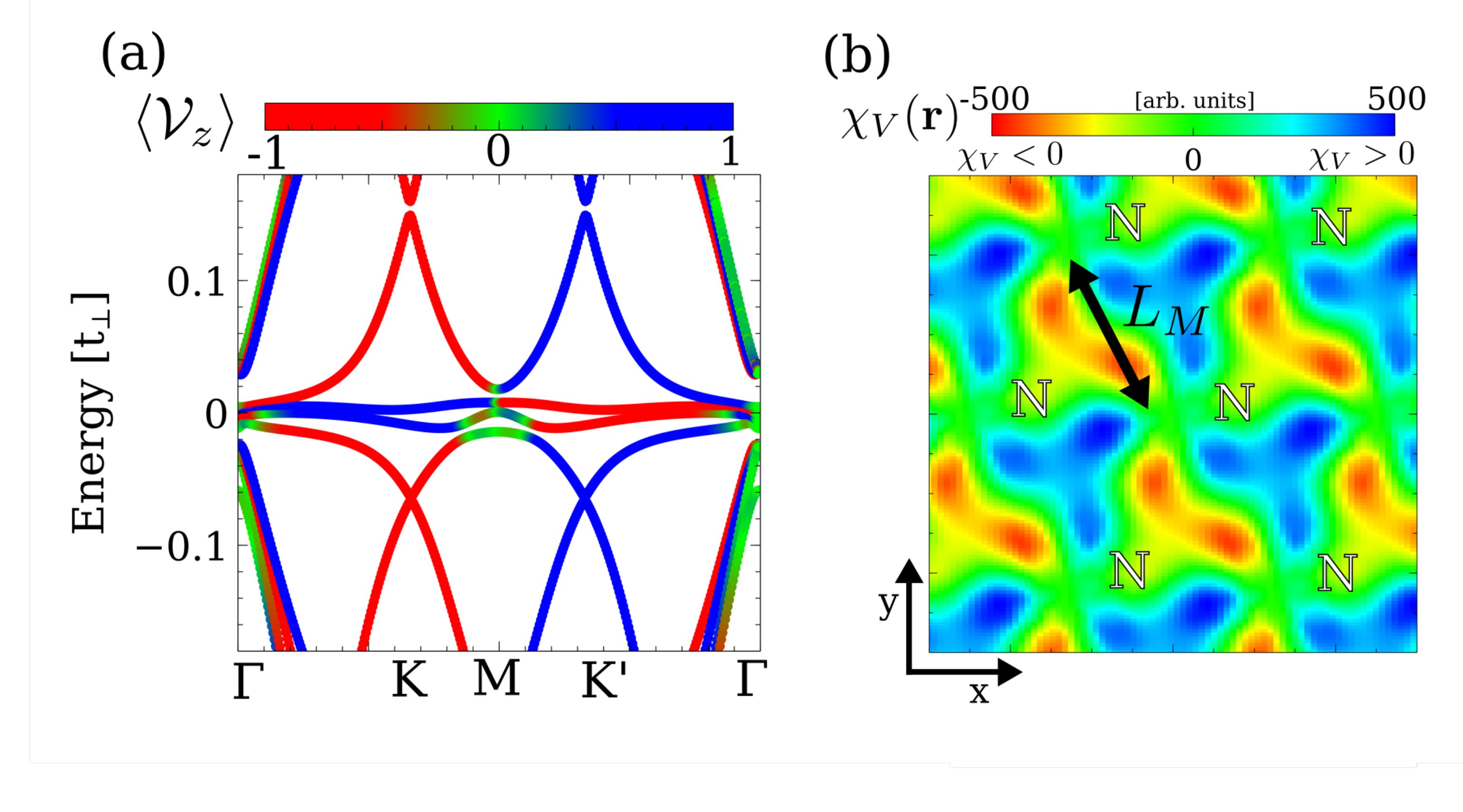}

\caption{
(a) Band structure in the presence of a N impurity, and (b) valley current at charge neutrality.
It is observed that, although the presence of a dopand creates a non-zero intervalley scattering,
the low-energy bands remain relatively flat. The emergence of intervalley scattering can be
seen in the depletion of the bulk valley currents in (b).
The valley flux in (b) is summed over the three layers.
Calculations performed with tight-binding.
}
\label{fig:doptb}
\end{figure}

We now consider the effect of a chemical 
impurity in the tight-binding model. We model the addition
of a chemical impurity by adding to the Hamiltonian $\mathcal{H}_D = w \sum_s c^\dagger_{i,s} c_{i,s}$,
where $i$ in the site that has been chemically replaced. 
We take $w=-2t$, which is the typical energy scale expected for a $N$
dopand, and yields results comparable with the first principle calculations.
For the sake of concreteness we will focus on the case with interlayer bias,
so our full Hamiltonian will be $\mathcal{H} = \mathcal{H}_0 + \mathcal{H}_V + \mathcal{H}_D$.
With the previous Hamiltonian,
we now compute the electronic band structure and project each eigenstate onto the valley operator.
The result is shown in Fig. \ref{fig:doptb}a, where we see that small anticrossings
appear as in the first-principles calculations. The valley projection
clearly shows that such anticrossings are
associated with intervalley mixing. The effect of the impurity in terms of intervalley mixing can
also be readily seen in the bulk valley currents. In particular, the existence of the impurity
is expected to strongly perturb the original valley fluxes in the unit cell,
depleting the local value of the valley Chern number.
This is verified
in Fig. \ref{fig:doptb}b, where we observe that the original valley fluxes are impacted
by the presence of the impurity.

Interestingly, despite the effect in terms of intervalley mixing, the low-energy bands
remain relatively flat, retaining their associated large density of states. This suggests
that correlated states can still appear in chemical doped twisted trilayers. It is worth
to emphasize that despite this large density of states, correlated phases relying on
valley coherent states will be strongly suppressed due to the impurity-induced
intervalley mixing. In particular, valley ferromagnet states, and valley triplet
superconducting states will be depleted due to chemical dopands. Nevertheless,
conventional spin-singlet valley-singlet states are not affected by the presence
of intervalley scattering. Motivated by this, in the next section we address the
emergence of spin/valley singlet superconductivity, and show
how the superfluid density impacts the high energy dispersive states.

\section{Superconducting state}
\label{sec:sc}

The large density of states close to charge neutrality suggests that the
twisted graphene trilayer can have superconducting instabilities,
similarly to twisted bilayers and tetra-layers. 
As shown above, both with electrostatic and chemical doping the
system shows divergent density of state close to charge neutrality.
For the sake of concreteness we will now focus on the
electrostatically doped system, yet we have verified that our results
remain qualitatively unchanged with chemical doping.

An emergent superconducting state
is associated with a Fermi surface instability, yet its effect can give rise
to second order perturbations above the Fermi energy.
In order to understand the potential impact at high-energies,
we first briefly analyze the structure of the low-energy states.
This can be done by comparing the spatial distribution of the low-energy states
with respect to the Fermi surface states. In particular, we define the projection
over the Fermi surface states as
$
    \mathcal{R} = \int d^2 \mathbf r |\Psi(\mathbf r)|^2 \rho_F (\mathbf r)
$
with $\rho_F (\mathbf r)$ the local density at the Fermi energy
$\rho_F (\mathbf r) \sim \int \langle \mathbf r | \delta (E_F - H) | \mathbf r \rangle$.
The quantity $\mathcal{R}$ allows to qualitatively distinguish which states are localized 
in the same region at the Fermi surface states.

\begin{figure}[t!]
\centering
    \includegraphics[width=\columnwidth]{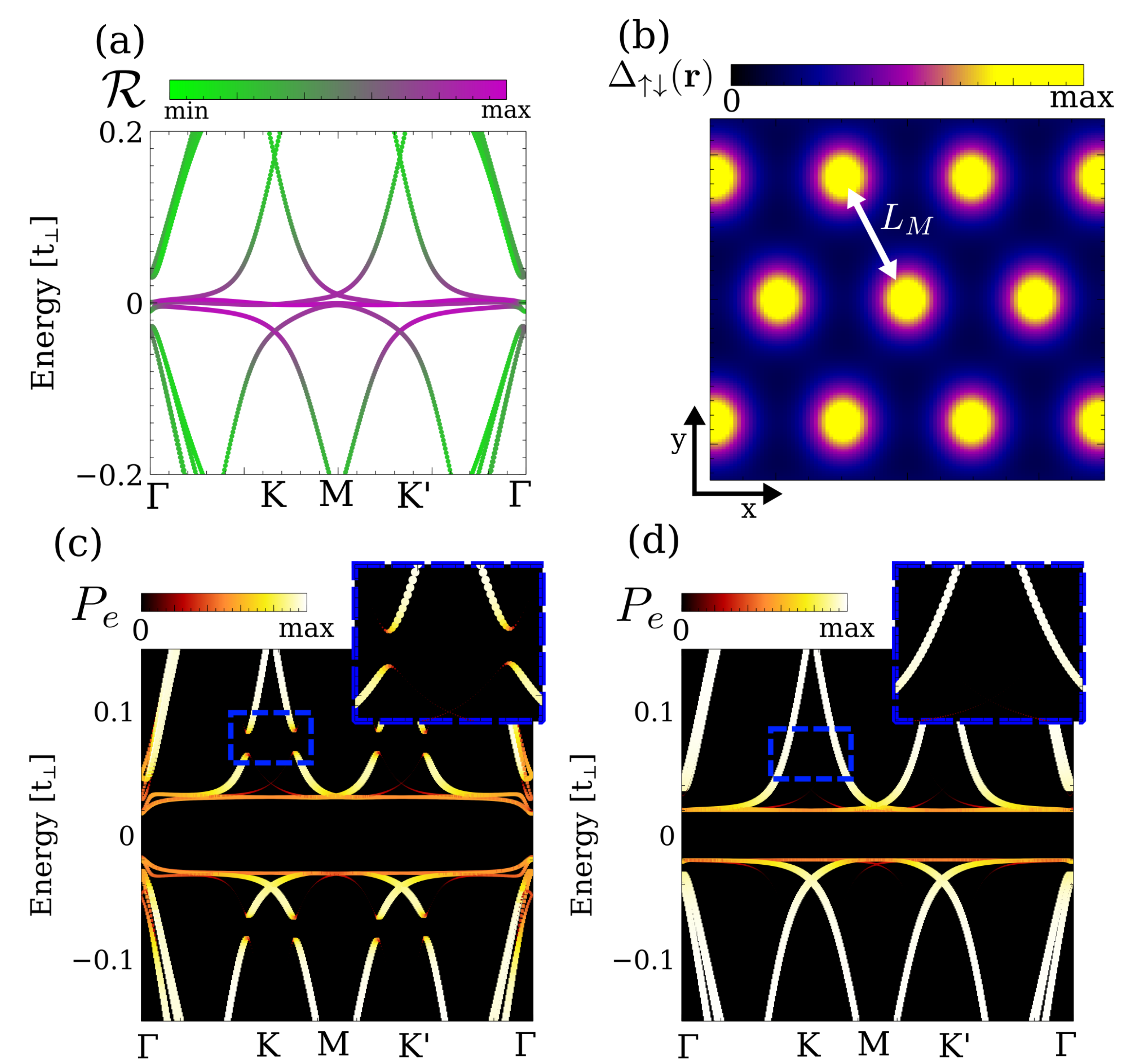}

\caption{
	Normal state band structure (a) in the absence of interactions,
	with the color denoting
	the projection of the states on the density
	of states at the Fermi energy.
	Upon introducing interactions, 
	a non-zero superfluid density
	appears as shown in 
	panel (b).
	Panel (c) shows the associated
	electronic dispersion (c) to the selfconsistent
	solution of (b), where
	we have projected the eigenstates over the
	the electron sector. Panel
	(d) shows the electronic dispersion
	projected in the electron sector
	for a uniform
	superfluid density, highlighting
	that the high energy anticrossings stem
	from the non-uniform pairing.
	Calculations performed with tight-binding.
}
\label{fig:figsc}
\end{figure}

By computing the Fermi surface state projector $\mathcal{R}$, it observed
that the states of the flat band are localized in similar regions.
In contrast, as one departs from the
Fermi energy, the states start to delocalize to other regions of the unit
cell (Fig. \ref{fig:figsc}a). This highlights the different orbital
nature of the nearly flat bands (in purple) and low-energy
dispersive bands (in green). Importantly, and in strike 
contrast with twisted graphene
bilayers, the flat bands of this system are not decoupled from the dispersive
states, suggesting that the superconducting states of this system will have
a genuine multiorbital nature.

Given the unavoidable entanglement between the flat and dispersive bands, 
an effective model description of this twisted trilayer
cannot be easily performed. Therefore, in the following
we will study the emergent superconducting state by exploiting the
full atomistic model, including all the bands in our calculation.
A variety of mechanisms
have been suggested to give rise to attractive interactions in these
systems, including phonon,\cite{PhysRevB.98.220504,PhysRevLett.121.257001,PhysRevLett.122.257002} 
Coulomb\cite{PhysRevLett.122.026801} 
and magnon fluctuations.\cite{PhysRevLett.121.087001,PhysRevLett.121.217001}
For the sake of concreteness, we will focus on
effective local attractive interactions,
as employed in other twisted graphene systems
\cite{PhysRevLett.97.230404,PhysRevLett.98.146801,PhysRevLett.100.246808,PhysRevB.101.060505,Heikkil2019,PhysRevB.98.220504,PhysRevLett.121.257001,PhysRevLett.123.237002}

\begin{equation}
\mathcal{H}_I = - \sum_i
	g c_{i,\uparrow}^\dagger c_{i,\uparrow}
	c_{i,\downarrow}^\dagger c_{i,\downarrow}
\end{equation}

that we solve at the mean field level
$
	\mathcal{H}_I^{MF} = -g\sum_i
	\langle c_{i,\uparrow}^\dagger 
	c_{i,\downarrow}^\dagger \rangle
	c_{i,\uparrow}
	c_{i,\downarrow}
	+ h.c.
$
where $\langle c_{i,\uparrow}^\dagger 
	c_{i,\downarrow}^\dagger \rangle$ is computed self-consistently
	for the full Hamiltonian
	$\mathcal{H} = \mathcal{H}_0 + \mathcal{H}_V + \mathcal{H}^{MF}_I$,
	that we solve
	with the Bogoliubov-de Gennes (BdG) formalism.
The local attractive interaction $g$ will give rise to
a net superfluid density, with a moir{\'e} momentum structure of
$s$-wave symmetry. 
In the following we take $g=t$, and we verified
that our results remain qualitatively similar with
smaller interaction strengths.
We note that although interactions are local,
the would lead to a non-trivial multiorbital structural in the moir{\'e}
orbital space.

By solving the previous selfconsistent problem, we find that the
superfluid density is non-uniform in the moir{\'e} unit cell
(Fig. \ref{fig:figsc}b), stemming from the
non-uniformity of the low-energy states.
By projecting the BdG eigenstates in the electron sector
via the electron projector $P_e$,
we observe
that the net non-uniform superfluid density 
$\Delta_{\uparrow\downarrow} (\mathbf{r}_i) \sim \langle c^\dagger_{i,\uparrow} c^\dagger_{i,\downarrow} \rangle$
gives rise to a full gap in the Brillouin zone
in the superconducting state. This phenomenology is similar to the
one found in twisted graphene bilayers.
More interestingly, besides the gap opening at the
chemical potential, anticrossings appear in high energy bands when the
selfconsistent pairing is included (Fig. \ref{fig:figsc}c). It is worth
to emphasize that,
in the presence of a uniform pairing artificially imposed, the high
energy anticrossings disappear, leading only to the gap opening at charge
neutrality (Fig. \ref{fig:figsc}d).
The emergence of gap openings away from charge neutrality is associated to
the intrinsically multiorbital nature of the superconducting state, and
stems from the non-unitarity of the superconducting matrix
in orbital subspace.\cite{PhysRevResearch.1.033107,2019arXiv190711602Z}
Interestingly, this shows that signatures of the superconducting state
can be obtained by analyzing the system away from the chemical potential,
and could provide powerful spectroscopic
signatures\cite{PhysRevB.97.165414,2020arXiv200202289L} of the superconducting state. 

\section{Conclusions}
\label{sec:sum}

By combining first-principles calculation and low-energy effective models,
we have shown that twisted graphene trilayers realize tunable electronic systems.
In particular, is was shown that nearly perfect flat bands can be
electrically controlled, that coexist with highly dispersive
states. Interestingly, such 
electric flattening of the bands is accompanied by the emergence of
bulk valley currents. We have found both from first
principles and low-energy calculations that chemical doping
does not destroy the flat bands, yet it substantially impacts
the bulk valley currents. This suggests that chemical doping
of twisted graphene trilayers could provide an intrinsic way of providing
the necessary electronic doping required for the emergence of
a superconducting state. We finally demonstrated that 
an emergent superconducting state
would give rise to spectroscopic changes int he high energy bands,
associated to the non-uniform superfluid density. Our results
highlight the rich physics of twisted graphene trilayers,
and provide a starting point to explore the interplay between
flat bands, correlation and dispersive states in twisted
graphene multilayers.

\section*{Acknowledgments}
Los Alamos National Laboratory is managed by Triad National Security, LLC, for the National Nuclear Security Administration of the U.S. Department of Energy under Contract No. 89233218CNA000001. 
This work was supported by the U.S. DOE Office of Basic Energy Sciences Program (E3B5).
A.L.-B. acknowledges the computing resources provided on Bebop, the high-performance 
computing clusters operated by the Laboratory Computing Resource Center at Argonne 
National Laboratory. J.L.L. thanks T. Wolf, G. Blatter, M. Sigrist,
O. Zilberberg, W. Chen, T. Neupert, A. Ramires and T. Heikkil\"a
for fruitful discussions. J.L.L. acknowledges the computational 
resources provided by the Aalto Science-IT project.

\section*{Appendix}
\appendix

 \section{\label{sec:methods } First-principles calculations}
Description of the coupling between the graphene layers was conducted through self-consistent calculations with the SIESTA code\cite{Soler2002} within a localized orbital basis set scheme. Paramagnetic calculations were conducted using a double-$\zeta$ basis set, and the local density approximation (LDA) approach\cite{PhysRev.140.A1133} for the exchange-correlation functional was used. Atomic positions of systems formed by over 5514 atoms were fully relaxed with a force tolerance of 0.02 eV/\AA. The integration over the Brillouin zone (BZ) was performed using a Monkhorst sampling in $\Gamma$ point. The radial extension of the orbitals had a finite range with a kinetic energy cutoff of 50 meV. A vertical separation of 35 \AA\ in the simulation box prevents virtual periodic parallel layers from interacting.

\section{The valley operator}
\label{sec:valley}

\begin{figure}[t!]
\centering
    \includegraphics[width=\columnwidth]{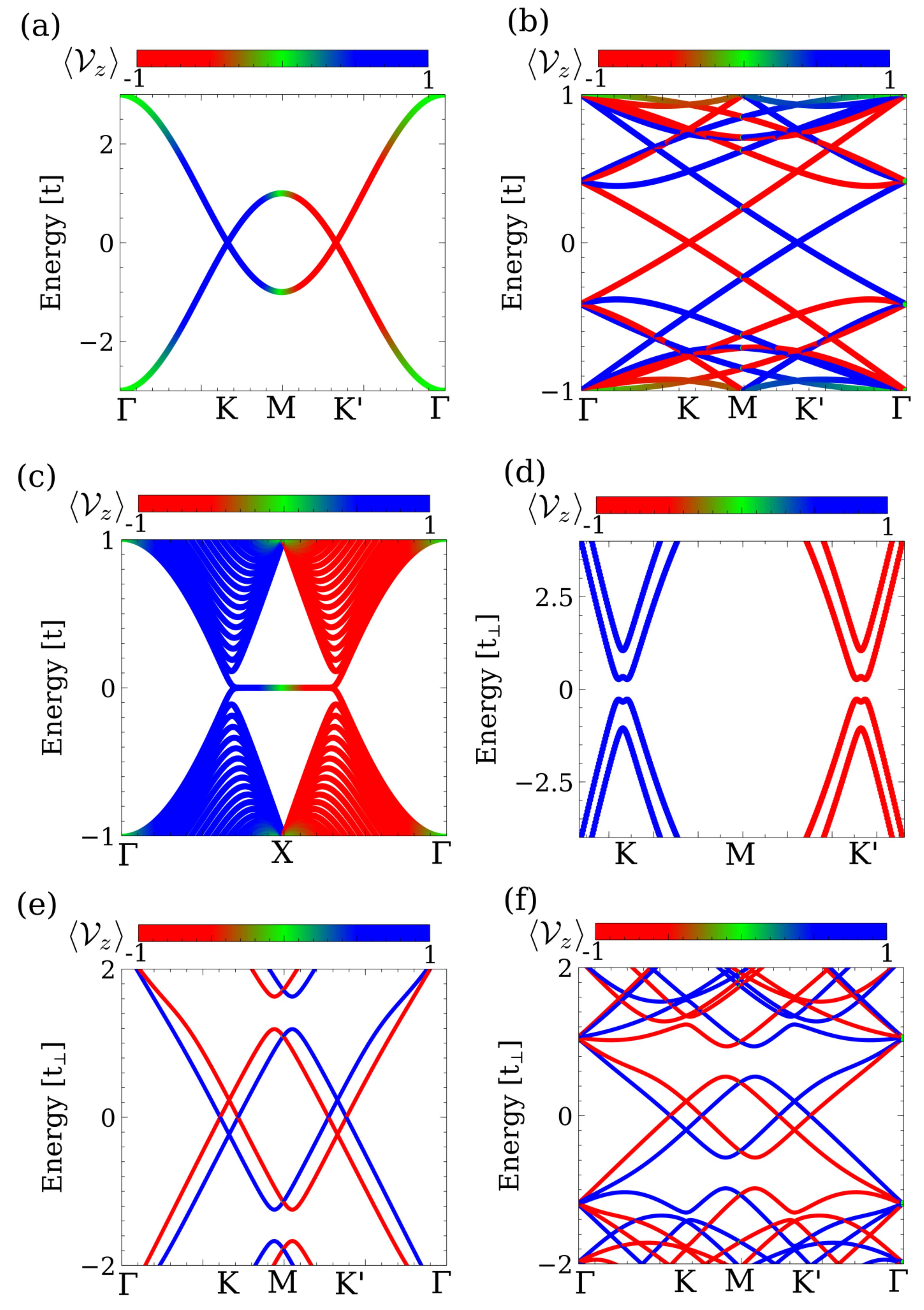}

        \caption{ Valley expectation
        value for the minimal unit cell
        of a honeycomb lattice (a) and in an 8x8 supercell.
        Panel (c) shows the projection of the valley operator
        on the states of a graphene zigzag ribbon,
        and panel (d) on a biased aligned Bernal stacked
        graphene bilayer.
        Panels (e,f) show the projection of the valley operator
        for a biased twisted graphene bilayer,
        for a rotation angle of 10$^\circ$ in (e) and
        5$^\circ$ in (f).
        We took $V=0.05t$ for (d,e,f).
}
\label{fig:SM2}
\end{figure}

Valley is an emergent quantum number in graphene, and as a result, it is
not apparent how a valley operator can be described in a real space 
tight-binding basis. A simple procedure to define the valley expectation value
in a tight binding model is by noting that for the z-component of the
valley, we are looking for an operator
with eigenvalue +1 for states in one valley, and -1 for states in another
valley. This would be accomplished by a Hamiltonian that realizes a
valley-dependent chemical potential. Let us first focus in a
honeycomb lattice, and let us take the following real space operator
in a honeycomb lattice\cite{PhysRevLett.120.086603,PhysRevLett.121.146801,PhysRevB.99.245118}

\begin{eqnarray}
        \mathcal{V}_z =\frac{i}{3\sqrt{3}}
        \sum_{\langle\langle i,j \rangle\rangle,s}
        \eta_{ij} \sigma_z^{ij} c^\dagger_{i,s} c_{j,s},
\label{eq:valley}
\end{eqnarray}

where $\langle\langle i,j \rangle\rangle$ denotes second neighbor sites,
$\eta_{ij}=\pm 1$ for clockwise or anticlockwise
hopping, and $\sigma^{ij}_z$ is a Pauli matrix associated with the sublattice
degree of freedom. The previous operator is diagonal
in sublattice, and is proportional
to the identity matrix in reciprocal space
$
\begin{pmatrix}
        f(\mathbf k) & 0 \\
        0 & f(\mathbf k) \\
\end{pmatrix}
$
with $f(\mathbf k)$ the Fourier transform of
the real-space hopping.
It is easily shown that
$f(\mathbf k) \approx +1$ close to the
K-point, and $f(\mathbf k) \approx -1$
close the K'-point, and as a result such
second neighbor hopping allows
to compute the expectation value of the valley for a specific
state. In particular, by taking the expectation value
$ \langle \mathcal{V}_z \rangle = 
\langle \Psi | \mathcal{V}_z | \Psi \rangle$, we will
obtain $ \langle \mathcal{V}_z \rangle \approx 1$ if $\Psi$ is a state belonging
to valley K, and $ \langle \mathcal{V}_z \rangle \approx -1$ if
$\Psi$ is a state belonging to valley K'. This can be clearly
seen in Fig. \ref{fig:SM2}a, where we show the valley expectation value
for the states of the honeycomb lattice, showing that the
states around valley K have eigenvalue +1, and around K' eigenvalue
-K'.

The valley operator allows us to easily track the valley flavor of the electronic
states in various situations. Let us now show some of them for the
sake of clarity. The simplest case consists of the electronic structure
of a supercell of a honeycomb lattice. In particular, we show in
Fig. \ref{fig:SM2}b the bandstructure for 8x8 supercell, indicating that the
valley operator allows following the original valley
flavor of the states in the folded band structure.
It is worth to note that such operator can be defined in a graphene
structure, without requiring two-dimensional
periodicity, for example for
graphene nanoribbons.
In particular, we show in \ref{fig:SM2}c the band-structure
of a zigzag graphene nanoribbon, demonstrating that the valley operator
correctly identifies the valley flavor of each state.

The valley operator can be easily extended to graphene multilayers. In
particular, by defining the valley operator of layer $\alpha$
as $\mathcal{V}^\alpha_z$, the total valley operator
is defined as
$\mathcal{V}_z = \sum_\alpha \mathcal{V}^\alpha_z$. Let us first
illustrate this with a simple graphene multilayer, an electrically
biased Bernal stacked
graphene bilayer. In this situation shown in
Fig. \ref{fig:SM2}d, we again observe that the
multilayer valley operator correctly identifies the states belonging
to the different valleys. This very same
idea can be used for twisted graphene multilayers. In particular,
in Fig. \ref{fig:SM2}ef we show that in a biased twisted graphene
bilayer at an angle of 10 degrees (Fig. \ref{fig:SM2}e) and
5 degrees (Fig. \ref{fig:SM2}f), the valley operator
correctly identifies the microscopic valley of each state.

 \section{Origin of the high energy anticrossings in the superconducting state}
 
 \begin{figure}[t!]
\centering
    \includegraphics[width=\columnwidth]{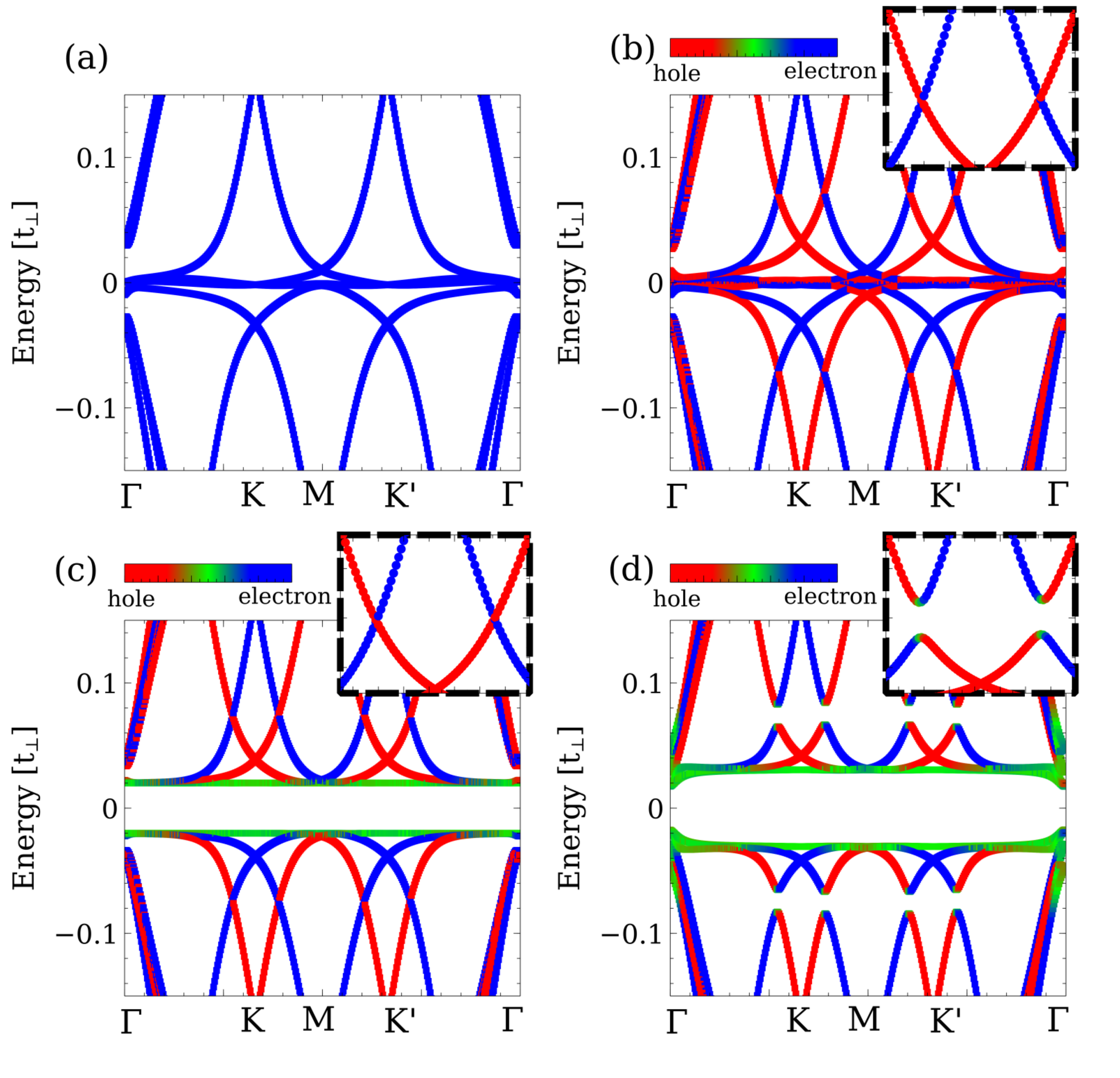}

        \caption{ Band structure of the biased twisted trilayer
        in the normal state, showing only the electron sector (a)
        and both the BdG electron and hole sectors (b).
        When a non-zero superfluid weight is included,
        anticrossings
        in the high energy bands do not appear for a uniform
        superfluid matrix $\Delta \sim \mathcal{I}$ (c). In contrast,
        anticrossings at high energies emerge
        for a self-consistent
        non-uniform superfluid weight (d), due to the
        non-zero matrix element
        $\Gamma = \langle \Psi_{n,\mathbf {k}}|\hat
\Delta | \Psi_{l,\mathbf {-k}} \rangle \ne 0$
}
\label{fig:SM1}
\end{figure}

In the following, we elaborate on the origin of the high energy anticrossings
in the superconducting state.
Let us start with the band structure of the biased twisted trilayer graphene,
as shown in Fig. \ref{fig:SM1}a.
By extending the spectra in a Bogoliubov de Gennes formalism,
hole replicas of the original states appear.
This
is shown in Fig. \ref{fig:SM1}b for $\Delta=0$, where the blue bands
denote electron-like bands, and red bands denote the hole replicas.
It is worth to note that in this situation, electron-like and hole-like
bands above the chemical potential cross, and therefore a non-zero superfluid
weight can potentially lead to anticrossings, as such
terms couples electron and hole sectors. Let us now turn on a spatially
uniform superfluid weight, as shown in Fig. \ref{fig:SM1}c. In this situation,
it is observed that for the electron and hole-like bands
cross above the
chemical potential no anticrossings appears. The coupling
between the electron and hole states above the chemical
potential is proportional to the overlap of the single
particle wavefunction $\Psi_{n,\mathbf{k}}$
with the superconducting matrix $\hat \Delta$, and takes the form
$\Gamma = \langle \Psi_{n,\mathbf {k}}|\hat
\Delta | \Psi_{l,\mathbf {-k}} \rangle $
with $n\ne l$.
For uniform superfluid density $\hat \Delta \sim \mathcal{I}$ with
$\mathcal{I}$ the identity matrix, we
have $\Gamma=0$ from the orthogonality of the wavefunctions, and therefore
no anticrossings appear in the high energy bands of Fig. \ref{fig:SM1}c.
In stark contrast, when the
superfluid weight is non-uniform in space
we have $\hat \Delta \not\propto \mathcal{I}$, we generically have
$\Gamma = \langle \Psi_{n,\mathbf {k}}|\hat
\Delta | \Psi_{l,\mathbf {-k}} \rangle \ne 0$,
leading to an effective anticrossing
between the states. As a result, the appearance of anticrossing
in the high energy bands is a direct consequence of the non-uniform
superfluid weight. We finally note that this argument
relies on the original time-reversal symmetry of the twisted
trilayer graphene Hamiltonian.

\bibliographystyle{apsrev4-1}
\bibliography{biblio}{}
\end{document}